\documentclass[reprint,aps,showpacs,pra,nofootinbib,superscriptaddress]{revtex4-2}
\usepackage{color}
\usepackage{newlfont}
\usepackage{graphicx}
\usepackage{amssymb}
\usepackage{amsmath}
\usepackage[caption=false]{subfig}
  
\usepackage{bm}
\usepackage{verbatim}
\usepackage[latin1]{inputenc}
\usepackage{bbm}
\usepackage{multirow}
\usepackage[thinlines]{easytable}
\usepackage{braket}
\usepackage{amsthm}
\usepackage{bbold}
\usepackage{subfig}
\usepackage{mathrsfs}
\usepackage[varg]{txfonts} 
\usepackage{amsfonts}

\newcommand{\ba}{\begin{array}}
\newcommand{\ea}{\end{array}}
\newcommand{\tx}{\text}
\newcommand{\mc}{\mathcal}
\newcommand{\mf}{\mathfrak}

\newcommand{\mr}{\mathrm}
\newcommand{\mbb}{\mathbbm}

\newcommand{\dt}{{\Delta t}}
\newcommand{\sch}{^\mathrm{\mf{s}}}
\newcommand{\zeroo}{^\text{\tiny $0$}}
\newcommand{\cm}{_\mathrm{\text{\tiny cm}}}
\newcommand{\rr}{_\mathrm{\text{\tiny r}}}
\newcommand{\sfrac}[2]{\text{\small $\frac{#1}{#2}$}}
\newcommand{\supdex}[2]{{#1}^\text{\tiny $#2$}}

\DeclareMathOperator{\Tr}{Tr}
\newtheorem{result}{Result}

\begin{document}

\title{Fluctuation theorems for genuine quantum mechanical regimes}
\author{T. A. B. Pinto Silva}
\email{pinto\_silva@campus.technion.ac.il}
\affiliation{Department of Physics, Federal University of Paran\'a, P.O. Box 19044, 81531-980, Curitiba, Paran\'a, Brazil}
\affiliation{Schulich Faculty of Chemistry and Helen Diller Quantum Center, Technion-Israel Institute of Technology, Haifa 3200003, Israel}
\author{R. M. Angelo}
\affiliation{Department of Physics, Federal University of Paran\'a, P.O. Box 19044, 81531-980, Curitiba, Paran\'a, Brazil}

\begin{abstract} 
Of indisputable relevance for non-equilibrium thermodynamics, fluctuations theorems have been generalized to the framework of quantum thermodynamics, with the notion of work playing a key role in such contexts. The typical approach consists of treating work as a stochastic variable and the acting system as an eminently classical device with a deterministic dynamics. Inspired by technological advances in the field of quantum machines, here we look for corrections to work fluctuations theorems when the acting system is allowed to enter the quantum domain. This entails including the acting system in the dynamics and letting it share a nonclassical state with the system acted upon. Moreover, favoring a mechanical perspective to this program, we employ a concept of work observable. For simplicity, we choose as theoretical platform the autonomous dynamics of a two-particle system with an elastic coupling. For some specific processes, we derive several fluctuation theorems within both the quantum and classical statistical arenas. In the quantum results, we find that, along with entanglement and quantum coherence, aspects of inertia also play a significant role since they regulate the route to mechanical equilibrium. 
\end{abstract}

\maketitle

\section{Introduction}

Fluctuation theorems (FTs) have been extended from the field of stochastic thermodynamics to general quantum scenarios~\cite{Hanggi2017,Gemmer2005,Deffner2019,Millen2016,Campisi2011,Binder2018,Talkner2016}, standing out as insightful relations connecting fluctuating quantities to aspects of thermal equilibrium. Among the regarded fluctuating quantities, work plays a prominent role for two reasons: (i) it is of key relevance for complete statements of the energy conservation law and (ii) work FTs yield sensible formulations of the second law of thermodynamics in terms of equilibrium free energy.

Paramount for any unambiguous definition of work is the specification of both the ``acting system'' (from now on referred to as \emph{agent}), the one that applies the driving force, and the ``system acted upon'' (hereafter, \emph{receiver}), the one which the force is applied on. Of course, in light of Newton's third law, there is no fundamental reason preventing one to assign to a given physical system the role of either agent or receiver (this labeling is done by free choice), but the notion of work and internal energy can only make sense through such a clear definition.

\begin{figure}[tb]
\centerline{\includegraphics[scale=0.13]{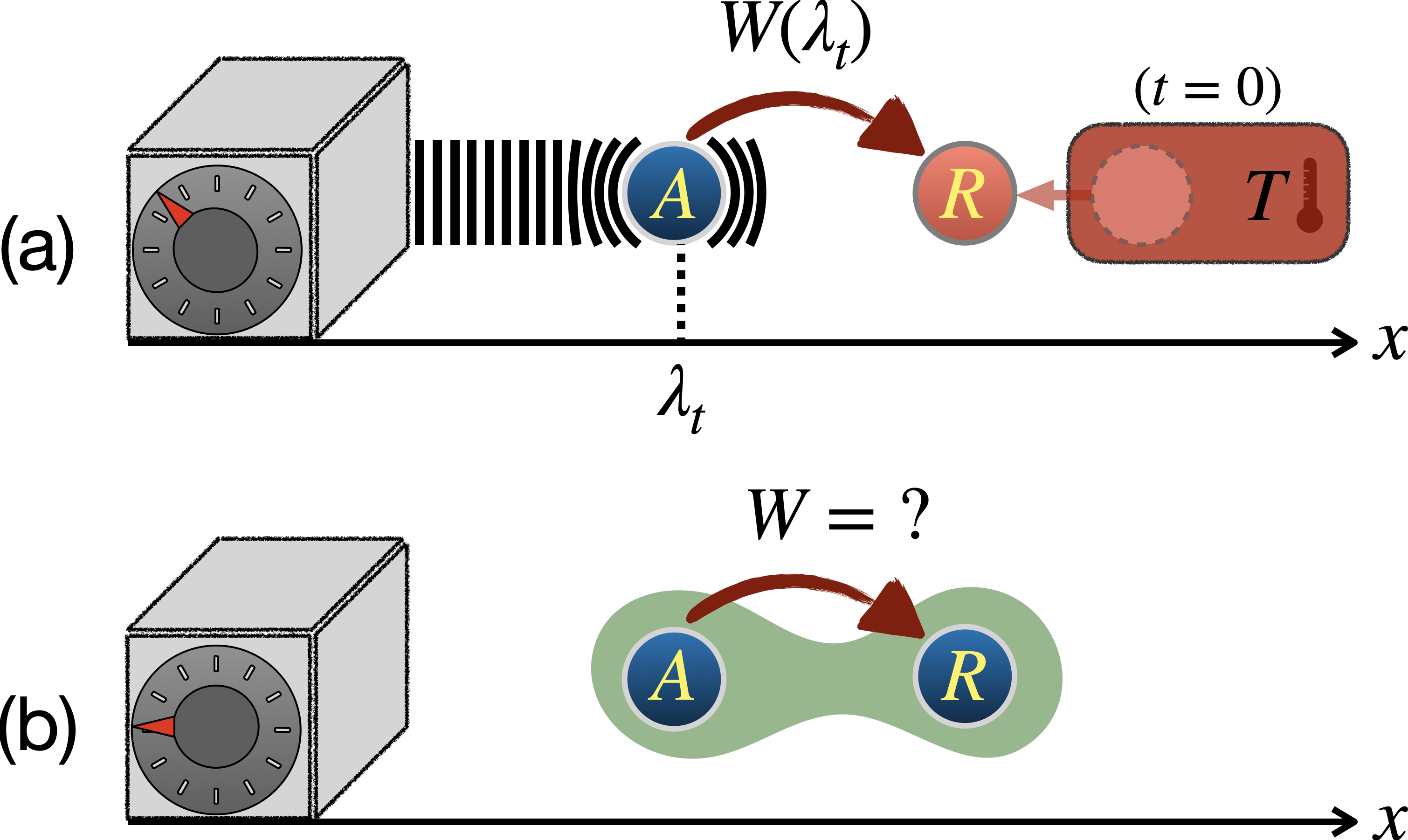}}
\caption{\small (a) Stochastic thermodynamics non-autonomous scenario. In the usual setting, a receiver $R$ is prepared at $t=0$ in a thermal state fully uncorrelated with the state of an agent $A$. For $t>0$, $R$ stops interacting with the thermal bath (of temperature $T$) and interacts only with $A$, whose position $\lambda_t$ is deterministically controlled by an external device. As a result, $A$ does positive work on $R$, and the usual forms of work FTs [e.g., Eqs.~\eqref{Jarintro} and \eqref{BKintro}] are obtained. (b)~ Quantum autonomous scenario. The controlling device is turned off and the (closed) system $A+R$ is abandoned to evolve autonomously, eventually via strong interactions and prepared in nonclassical states (possessing quantum coherence and quantum correlations). Again, $A$ does work on $R$, but now the underlying statistics and work FTs are entirely dictated by the rules of quantum mechanics, this being the case also for the very notion of work.}
\label{fig1}
\end{figure}

In the usual stochastic thermodynamics scenarios, the agent\footnote{In fact, there are some approaches in which the classicalization is introduced in the receiver's dynamics instead of the agent's~\cite{Peliti2008a}. Nonetheless, this is not the usual perspective.} is classical in essence, meaning that it is rigidly controlled by an external observer who assigns to it a pre-determined time dependence [see Fig.~\ref{fig1}(a)]. In effect, the agent's influence over time is entirely encoded in a function $\lambda_t\equiv \lambda(t)$~\cite{Campisi2011,Deffner2019,Sekimoto2010}, a prescription also adopted in the formalism of statistical
physics~\cite{Reif2009}. As a result, the Hamiltonian describing the receiver's dynamics is an explicitly time-dependent function usually written in the form $\mathcal{H}(t)\equiv \mathcal{H}(\lambda_t)=\mathcal{H}_{0}+V(\lambda_t)$, where $\mathcal{H}_{0}$ is the so-called bare Hamiltonian and $V(\lambda_t)$ is an interaction term~\cite{Jarzynski2007,Campisi2011}. Within this perspective, concepts of work and acclaimed work FTs~\cite{Campisi2011,Jarzynski1997, Bochkov1977, Crooks1999} were proposed, with work depending either implicitly or explicitly on $\lambda_t$~\cite{Jarzynski2007,Horowitz2007}.

Among the results known today, the Jarzynski equality~\cite{Jarzynski1997} certainly stands out, being suitable for a large set of applications and experimental platforms (see~\cite{liphardt2002,Campisi2011,Hanggi2017,Huber2008} and references therein), with some extensions to more general scenarios~\cite{Jarzynski2004,Campisi2009,Campisi2011}. Departing from the so-called \emph{inclusive work}~\cite{Jarzynski2007},  $\mathcal{W}_{\mr{inc}}(t,0)=\int_{0}^{t}dt'\frac{\partial \mathcal{H}(t')}{\partial t'}$, Jarzynski arrived at 
\begin{equation}
\braket{e^{-\beta \mathcal{W}_{\mr{inc}}(t,0)}} =\frac{\mathcal{Z}_{t}}{\mathcal{Z}_{0}},\label{Jarintro}
\end{equation}
with $\mathcal{Z}_{t}=\int \!d\Gamma\,e^{-\beta \mathcal{H}(t)}$ denoting the partition function at the instant $t$ and $d\Gamma$ the infinitesimal phase-space volume accessible to the receiver. In another vein, it was only posteriorly acknowledged~\cite{Campisi2011,Campisi2011a,Jarzynski2007,Horowitz2007} that a distinct fundamental relation had already been derived by Bochkov and Kuzovlev (BK) in their late 1970s article~\cite{Bochkov1977}. BK deduced the equality
\begin{equation}
\braket{e^{-\beta \mathcal{W}_{\mr {exc}}(t,0)}} =1 \label{BKintro}
\end{equation}
under assumptions very similar to those of Jarzynski's approach, except that an \emph{exclusive} form of work, $\mathcal{W}_{\mr{exc}}(t,0)=\int_{0}^{t}dt'\frac{d \mathcal{H}_{0}}{dt'}$, was used instead. The distinction between inclusive and exclusive work led not only to different FTs, as in Eqs.~\eqref{Jarintro} and \eqref{BKintro}, but also to different work-energy relations, this being the source of an intense debate~\cite{Vilar2008,Vilar2008a,Vilar2008b,Peliti2008,Horowitz2008,Jarzynski2007,Horowitz2007}. It turns out that the inclusive approach used to deduce the Jarzynski equality is close to a \emph{thermodynamical} picture of work \cite{Peliti2008a,Gibbs1902}, where the interaction energy with external bodies is accounted as part of the internal energy. On the other hand, the exclusive definition has its essential features very closely related to those of a \emph{mechanical} notion of work, as in the Newtonian description of massive point particles, where the internal energy does not generally encompass external degrees of freedom~\cite{Vilar2008}.

When going to a quantum regime, the usual route to describe work and work FTs has been to directly quantize the classical Hamiltonian $\mathcal{H}(t)$ into a time-dependent operator $H(t)\equiv H(\lambda_t)$ describing the internal energy, with the agent's influence being encoded in the control parameter $\lambda_t$. As a consequence, $\lambda_t$-dependent work definitions analogous to the classical ones were proposed \cite{Alicki1979,Allahverdyan2004,Allahverdyan2005,Talkner2007,Peliti2008a,Roncaglia2014,Ribeiro2016,Francica2017,Hummer2001,Perarnau2017,Strasberg2021}. Still in the nonautonomous context, a debate emerged around the fact that some of the work definitions lead to results that deviate from the usual classical FTs~\cite{Allahverdyan2004,Allahverdyan2005,Talkner2007,Allahverdyan2014,Engel2007}. Indeed, it was later shown that work definitions could not simultaneously satisfy two natural requirements, namely, (i) that mean energy variation corresponds to average work and (ii) that work statistics agree with usual classical results for initial states with no coherence in the energy basis~\cite{Perarnau2017}. As a result, further work definitions were introduced and modified work FTs deduced~\cite{Campisi2011,Deffner2016,Allahverdyan2014,Allahverdyan2005}. In none of these approaches, however, the role of the agent's configuration was critically addressed from a purely quantum substratum, so that the use of a classical parameter $\lambda_t$ was unavoidable.

Now, what if the external control is turned off and the composite system ``agent $+$ receiver'' is let to evolve autonomously, as depicted in Fig.~\ref{fig1}(b)? How can one compute and make sense of work FTs? With the ever-growing interest in thermodynamics phenomena in the quantum regime, the search for generalizations of the concept of work and FTs for autonomous scenarios started to make sense. Within this agenda, autonomous machines have been analyzed~\cite{Tonner2005,Lorch2018,Gelbwaser2013,Gelbwaser2015}, the effects of correlations, coherences, degeneracy, thermal fluctuations, and information resources on thermodynamics have been studied~\cite{Gelbwaser2014,Niedenzu2015,Weimer2008,Manzano2018,Alipour2016}, and batteries (or work reservoirs) have explicitly been considered in the dynamics~\cite{Skrzypczyk2014,Deffner2013}. Alternatively, some constraints (sometimes taken as general quantum FTs) have been obtained for a general class of systems~\cite{Brandao2013,Aberg2018,Alhambra2016} and other forms to statistically describe work and other thermodynamics properties have been discussed~\cite{Lostaglio2018,Janovitch2022,Park2020,Francica2022a,Francica2022b}. Still, no detailed analysis has so far been reported describing how the well-known FTs are affected in dynamics far away from the usual stochastic thermodynamics regimes, in particular when the agent is no longer classical. Filling this gap is the primary goal of the present work.

Here, we examine how work FTs manifest themselves when we push the system a bit further into the quantum domain. Basically, we pursue a fundamentally mechanical treatment characterized by two key elements. First, we let the agent be submitted, along with the receiver, to a closed energy-conserving autonomous dynamics, upon which no external control $\lambda_t$ is ever imposed. In particular, we allow the composite system ``agent $+$ receiver'' to be prepared in nontrivial quantum states, eventually encoding coherence, quantum correlations, and local thermal effects. Moreover, we allow the subsystems to strongly interact with each other without demanding the interaction to be time-independent or even to fade over time \cite{Brandao2013,Aberg2018,Alhambra2016}. To the best of our knowledge, these regimes have remained widely unexplored so far. Second, we abandon the usual thermodynamical essence assigned to work in favor of an operator-based model, which naturally attaches a fundamentally quantum mechanical character to this concept. In effect, this model treats work as a Heisenberg observable admitting an eigensystem for each given process and genuine quantum fluctuations~\cite{Silva2021}. Despite some skepticism to treating work as an observable \cite{Campisi2011,Talkner2007,Allahverdyan2014,Talkner2016}, the work observable formalism was shown to be experimentally testable and physically sound, besides being approachable as a two-time element of reality. At last, taking the operator $W$ as the work done by the agent on the receiver, we compute the average $\braket{e^{-\beta_1 W}}$, with $\beta_1$ being an effective inverse temperature underlying the receiver's initial state. To free the discussion of unnecessary technicalities, our theoretical platform is chosen to be as simple as possible: we consider a two-particle system, with elastic coupling, evolving over specific time intervals. Although the framework here introduced to evaluate $\braket{e^{-\beta_1 W}}$ can be applied to more general scenarios, this simple system suffices to show that work FTs may depend on the features of the receiver and agent. Our results are then compared with the BK equality~\eqref{BKintro}, which is closer to the mechanical paradigm than Jarzynski's formula~\eqref{Jarintro}. To highlight the genuinely quantum aspects of our results, we conduct classical studies in parallel employing the usual Newtonian notion of work, with its statistics being raised in accordance with the Liouvillian framework. Although our work FTs are shown to accurately retrieve BK's equality in some regimes (see also the Appendix for a related discussion), they manifest themselves rather differently, and somewhat surprisingly, in quantum instances.

\section{Classical autonomous scenario\label{CA Scenarios}}

We start by investigating the classical statistical framework, wherein the celebrated FTs were originally derived. Consider two particles of masses $m_{1,2}$ interacting via an elastic potential of characteristic constant $k$. The autonomous dynamics is governed by the Hamiltonian function
\begin{equation}
\mathcal{H}=\frac{p_{1}^{2}}{2m_{1}}+\frac{p_{2}^{2}}{2m_{2}}+\frac{k}{2}(x_{2}-x_{1})^{2},
\label{H-elasticcl}
\end{equation}
where $x_i$ ($p_i$) is the position (momentum) of the $i$-th particle. Henceforth, particle $1$ ($2$) will assume the role of receiver (agent). Within a mechanical perspective, the work $\mathcal{W}(t_2,t_1)$ done by particle $2$ on particle $1$ during the time interval $[t_1,t_2]$, is defined as
\begin{equation}
\mathcal{W}(t_2,t_1) \coloneqq\int_{t_1}^{t_2}\!\!dt\,\,m_1\ddot{x}_1\, \dot{x}_1
\displaystyle=\Delta\mathcal{K},
\label{RQMWcl} 
\end{equation}
where $\Delta\mathcal{K}\!=\!\mathcal{K}_1(t_2)\!-\!\mathcal{K}_1(t_1)$, with  $\mathcal{K}_1(t)\!=\!\frac{m_1\dot{x}_1^2(t)}{2}$ being the kinetic energy of particle $1$ (receiver's internal energy). Thus, Eq~\eqref{RQMWcl} is the usual statement of the energy-work theorem~\cite{Kleppner2010}. 

Aiming at computing the statistics underlying $\mathcal{W}(t_2,t_1)$, we explicitly solve the Hamilton equations in terms of the initial phase space point $\Gamma_{0}=(x_{1}\zeroo,p_{1}\zeroo,x_{2}\zeroo,p_{2}\zeroo)\equiv [x_{1}(0),p_{1}(0),x_{2}(0),p_{2}(0)]$. The procedure is facilitated by the use of the center-of-mass and relative coordinates
\begin{equation}
\begin{array}{lcl}
\displaystyle x\cm&=&(m_1x_1+m_2x_2)/M,\\ \displaystyle x\rr&=&x_2-x_1,\\ 
\displaystyle p\cm&=&p_1+p_2,\\
\displaystyle p\rr&=&\mu\left(p_2/m_2-p_1/m_1\right),
\end{array}\label{eq:relacmcl}  
\end{equation}
with $\mu=m_1m_2/M$ and $M=m_1+m_2$. In the transformed Hamiltonian,  $\mathcal{H}=p\cm^2/2M+p\rr^2/2\mu+kx\rr^2/2$, the new degrees of freedom decouple and the trajectories are trivially derived:
\begin{equation}
\begin{array}{lcl}
x\cm^t&=&x\cm\zeroo+p\cm\zeroo t/M, \\  x\rr^t&=&x\rr\zeroo\cos\left(\omega t\right)+(p\rr\zeroo/\mu\omega)\sin\left(\omega t\right), \\  
p\cm^t&=&p\cm\zeroo,	\\
p\rr^t&=&p\rr\zeroo\cos\left(\omega t\right)-\mu\omega  x\rr\zeroo\sin\left(\omega t\right),\label{clcmr}
\end{array}
\end{equation}
with $\omega=\sqrt{k/\mu}$. Returning to the original variables, we can write an expression for the momentum of particle $1$ at a generic time $t$,
\begin{equation}
p_1^t(\Gamma_{0})=a(t)\,p_1\zeroo+b(t)\,p_2\zeroo+c(t)\left(x_2\zeroo-x_1\zeroo\right),
\label{P1-elasticl}
\end{equation}
\begin{equation}
\left\{\begin{array}{l}
a(t)=\left[m_1+m_2\cos\left(\omega t\right)\right]/M,\\ 
b(t)=\left[1-\cos\left(\omega t\right)\right]m_1/M,\\ 
c(t)=\mu\omega\sin\left(\omega t\right).
\end{array}\right.
\label{abc}
\end{equation}
As a result, we are able to write the kinetic energy $\mathcal{K}_{1}(t)$ and the work $\mathcal{W}(t_2,t_1)$ as explicit functions of the initial phase point $\Gamma_{0}$. For the sake of analytical convenience (which will be specially welcome within the quantum context treated posteriorly), hereafter we restrict our analysis to processes occurring within the time intervals $[t_1,t_2]=[0,v\tau]$, with $v$ an odd integer and $\tau=\pi/\omega$. With the notation $\mathcal{W}(\Gamma_{0})\equiv \mathcal{W}\left(v\tau,0\right)$, the resulting work can be written as
\begin{equation}
\mathcal{W}(\Gamma_0)=\frac{2}{M^2}(m_1p_2\zeroo-m_2p_1\zeroo)(p_1\zeroo+p_2\zeroo).
\label{W21-elasticl}
\end{equation}
To raise the work statistics, we consider an initial distribution $\varrho(\Gamma_0)$, so that the mean value of a well-behaved function $ f[\mathcal{W}(\Gamma_0)]$ is given by
\begin{equation}
\braket{f(\mathcal{W})}_\varrho=\int\!d\Gamma_0\, f[\mathcal{W}(\Gamma_0)]\,\varrho(\Gamma_0).\label{fwcl}
\end{equation}

\subsection{Case studies}
Focusing on scenarios associated with the BK equality, we consider, as our first case study, the initial thermal-Gaussian (TG) distribution
\begin{equation}
\varrho_\tx{\tiny TG}(\Gamma_0)=\mathcal{T}_{\Delta_1}(x_{1}\zeroo,p_{1}\zeroo)\,\, \mc{G}_{\bar{\mathbf{r}}_2,\sigma_2}(x_{2}\zeroo,p_{2}\zeroo),\label{varthg}
\end{equation}
which assigns to the receiver the thermal distribution 
\begin{equation}
\mathcal{T}_\Delta(x,p)\coloneqq\frac{\exp\left(-\frac{ p^2}{2\Delta^2}\right)}{\sqrt{2\pi\Delta ^2}}\,\varrho(x),\label{varrth}
\end{equation}
where $\Delta \coloneqq \sqrt{m/\beta}$ is a ``thermal momentum uncertainty'' (which also is an indirect measure of temperature), $\beta$ is an inverse temperature, and $\varrho(x)$ is a generic probability distribution\footnote{In some thermodynamic instances, $\varrho(x)$ was chosen to characterize a particle confined in a box~\cite{Pathria1996,Reif2009}.}. By its turn, the agent is given the Gaussian distribution $\mathcal{G}_{\bar{\mathbf{r}},\sigma}(x,p)\equiv\mathcal{G}_{\bar{x},\sigma_x}(x)\,\mathcal{G}_{\bar{p},\sigma}(p)$, with $\bar{\mathbf{r}}=(\bar{x},\bar{p})$, $\sigma_x=\hbar/(2\sigma)$, and
\begin{equation}
\mc{G}_{\bar{u},\sigma_u}(u)\coloneqq\frac{\exp\left[-\frac{(u-\bar{u})^2}{2\sigma_u^2}\right]}{\sqrt{2\pi\sigma_u^2}},\label{Gausscl}
\end{equation}
where $\bar{u}$ and $\sigma_u$ respectively denote the center and the width of the Gaussian distribution. By use of the distribution~\eqref{varthg}, the averaging prescribed by Eq.~\eqref{fwcl} for the function $f[\mathcal{W}(\Gamma_{0})]=e^{-\beta \mathcal{W}(\Gamma_{0})}$ results in
\begin{equation}
\braket{e^{-\beta_1 \mathcal{W}}}_{\varrho_\tx{\tiny TG}}=\frac{m_{1}+m_{2}}{|m_1-m_2|}.\label{FTclthg}
\end{equation}
In comparison with the BK formula~\eqref{BKintro}, the differences are clear and insightful, specially with regard to the finite inertia of the agent. Notably, the BK formula is recovered as $\frac{m_1}{m_2}\to 0$, the limit in which $x_2^t\to x\cm^t=x\cm\zeroo+p\cm\zeroo t/m_2$.  This is the precise regime for which the BK equality was deduced, viz. the one presuming that the agent acts as a deterministic classical driver whose dynamics (not necessarily uniform) can in no way be disturbed by the receiver. Therefore, the dependence of the result~ \eqref{FTclthg} on the masses is a direct consequence of the autonomous character of the dynamics under scrutiny. Also noticeable is the fact that Eq.~\eqref{FTclthg} does not depend on the details of the preparation, such as $\bar{\mathbf{r}}_2$, $\sigma_2$, and $\Delta_1=\sqrt{m_1/\beta_1}$. This particular aspect may be a consequence of the quadratic structure of the model and eventual peculiarities underlying the chosen time interval. In any case, this reveals that, as long as the condition $m_2\gg m_1$ is satisfied, the BK equality holds even in the regime of a highly fluctuating agent distribution, for which the notion of a deterministic classical control can no longer be sustained. In fact, as discussed in the Appendix, this is a special case of a more general result for autonomous dynamics: the BK equality can actually be retrieved whenever the agent's dynamics is sufficiently independent of the receiver's degrees of freedom. 

We now conduct our second case study, wherein both receiver and agent are initially given thermal states with respective inverse temperatures $\beta_1$ and $\beta_2$. The composite thermal-thermal (TT) distribution reads
\begin{equation}
\varrho_\tx{\tiny TT}(\Gamma_0)=\mathcal{T}_{\Delta_1}(x_{1}\zeroo,p_{1}\zeroo)\,\,\mathcal{T}_{\Delta_2}(x_{2}\zeroo,p_{2}\zeroo),\label{varthth}
\end{equation}
where $\Delta_i=\sqrt{m_i/\beta_i}$ with $i\in\{1,2\}$. The calculations show that the result is identical to the previous one, that is,  $\braket{e^{-\beta_1 \mathcal{W}}}_{\varrho_\tx{\tiny TT}}=\braket{e^{-\beta_1 \mathcal{W}}}_{\varrho_\tx{\tiny TG}}$. Again, the inertial aspects are seen to prevail over any other elements of the preparation, even the arbitrary temperatures $\beta_{1,2}$.

The situation gets more interesting when we come to our third case study. Here we let not only thermal ingredients be present but also correlations. The initial distribution is chosen to be
\begin{equation}
\varrho_c(\Gamma_0)=\mathcal{T}_{\Delta _1}(x_1\zeroo,p_1\zeroo)\,\, \varrho_{\text{\tiny {$x$}}}(x_2\zeroo)\delta_{\tx{\tiny D}}(p_2\zeroo - c p_1\zeroo),\label{varcorrth}
\end{equation}
where $\delta_{\tx{\tiny D}}$ is the Dirac delta function, $c\in\mathbbm{R}_{\geq 0}$ is a dimensionless parameter whose role is discussed below, and again $\varrho_{\text{\tiny {$x$}}}(x_2\zeroo)$ is a generic probability distribution. It is not difficult to check that both marginals are thermal distributions, that is, 
\begin{equation}
\!\!\varrho_i(x_i\zeroo,p_i\zeroo)\equiv\!\iint\!dx_j\zeroo\, dp_j\zeroo\,\varrho_c(\Gamma_0)=\mathcal{T}_{\Delta_i}(x_i\zeroo,p_i\zeroo),
\end{equation}
with $i,j\in\{1,2\}$ and $j\neq i$. An interesting aspect is the appearance of the local inverse temperature $\beta_{2}(c)\equiv\frac{\beta_1}{c^2}\frac{m_2}{m_1}$ deriving from the connection $\Delta_2=c\Delta_1$. We see, therefore, that $c$ is a direct estimate of both the correlations between the particles and the agent's thermal momentum uncertainty. Through the procedure established previously, we arrive at
\begin{equation}
\braket{e^{-\beta_1 \mathcal{W}}}_{\varrho_c}=\frac{m_{1}+m_{2}}{|m_{1}-m_{2}+2m_{1}c|}.\label{FTclcorr}
\end{equation}
In direct comparison with Eq.~\eqref{FTclthg}, the above result demonstrates that classical correlations can influence work FTs in a relevant way. In particular, for $m_1=m_2$, one has $\braket{e^{-\beta_1 \mathcal{W}}}_{\varrho_c}=\frac{1}{c}= \sqrt{\beta_2/\beta_1}$, which makes explicit the strong dependence of the result also on the local temperatures. From a broader perspective, result~\eqref{FTclcorr} reveals an interesting generalization of the BK formula: by getting apart from the typical thermodynamics setting wherein the agent is a deterministic driver, we find that work FTs can strongly depend on both inertia and, via correlations, agent's effective temperature, $T_2=[k_B\beta_2(c)]^{-1}$.  

With the aim of getting more insight about our results, it is opportune to make some digression on energetics. Direct application of Jensen's inequality, $\braket{g(X)}\geqslant g(\braket{X})$, with $g$ a convex function and $X$ a random variable, allows us to express the Jarzynski equality~\eqref{Jarintro} in the form $\braket{\mathcal{W}_\tx{inc}(t,0)}\geq \Delta F$, where $F_\mathfrak{t}=-\beta^{-1}\ln\mathcal{Z}_\mathfrak{t}$ denotes the equilibrium free energy at instant $\mathfrak{t}$ and $\Delta F=F_\mathfrak{t}-F_0$. This inequality bounds the mean inclusive work with quantities directly associated with the thermodynamic equilibrium and allows one to make inferences, through the sign of $\Delta F$, about the spontaneity of a physical process. On the other hand, no symptom of thermodynamic equilibrium shows up straightforwardly in the BK equality~\eqref{BKintro}. Still, the derivation of this formula presumes important thermodynamic elements, namely, the preparation of a thermal state for the receiver and an external classical control. These aspects are crucial for a deeper understanding of the relation $\braket{\mathcal{W}_\tx{exc}(t,0)}\geq 0$ bounding the mean exclusive work. Basically, this inequality states that the agent can only deliver energy to the receiver. This can be explained via the following rationale. One, the thermal state imposes to the receiver a scenario of energetic minimization constrained to a certain temperature. Two, the agent has no need to consume energy from its interaction with the receiver because the agent's dynamics is deterministically pumped by an external control. Thus, the average result of such dynamics cannot be other than an increase of the receiver's internal energy. To make contact with this framework, we rephrase our results as 
\begin{equation}
\braket{\mathcal{W}}_\varrho\geq -\beta_1^{-1}\ln\braket{e^{-\beta_1\mathcal{W}}}_\varrho.
\label{C-LB}
\end{equation}
For the processes under scrutiny, the work FTs~\eqref{FTclthg} and \eqref{FTclcorr} have shown to be independent of thermodynamic equilibrium quantities, like $\beta_1$ and $\Delta_1$, and strongly dependent on inertia, which is the physical element capable of tuning the mechanical equilibrium. In effect, when $m_{1(2)}\gg m_{2(1)}$ one of the particles approximately remains in uniform motion (mechanical equilibrium). Although the BK formula is retrieved in this regime, the scenarios are still different, since in BK's approach the agent's motion, being dictated by $\lambda_t$, does not need to be uniform. On the other hand, whenever $\braket{e^{-\beta_1\mathcal{W}}}_\varrho\gg 1$, the work FTs largely deviate form the BK formula, and the lower bound in relation~\eqref{C-LB} can become significantly negative, meaning that the agent is now allowed to draw energy from the receiver. This regime is favored when $m_1\simeq m_2$, an instance in which energy exchange between agent and receiver is expected to be ubiquitous throughout the dynamics and, hence, the concept of mechanical equilibrium evaporates. Interestingly, we see by Eq.~\eqref{FTclcorr} that, even in the regime of mechanical equilibrium ($m_2\gg m_1$), an amount $c=\frac{(m_2-m_1)}{2m_1}$ of classical correlations is able to significantly disturb the directionally of the energetic flow typical of the BK scenario. Given the above, it is fair to conclude that the work FTs we thus far obtained make important connections with elements of mechanical (instead of thermodynamic) equilibrium. 

\section{Quantum autonomous scenario}
In full analogy with the classical model studied in the previous section, we now consider particles of masses $m_1$ and $m_2$ evolving autonomously under the unitary dynamics implied by the Hamiltonian operator
\begin{equation}
H=\frac{P_{1}^{2}}{2m_{1}}+\frac{P_{2}^{2}}{2m_{2}}+\frac{k}{2}(X_{2}-X_{1})^{2},
\label{H-elastic}
\end{equation}
where $X_i$ ($P_i$) is the position (momentum) operator of the $i$-th particle. The quantum preparation $\rho$ and $H$ act on the joint Hilbert space $\mbb{H}=\mbb{H}_1\otimes\mbb{H}_2$. As before, particle $1$ ($2$) will play the role of the receiver (agent).

Since we are interested in exploring FTs under a mechanical perspective, we employ the definition of work proposed in Ref.~\cite{Silva2021}.  Accordingly, we use the Heisenberg picture, wherein the operators evolve in time according to the relation $O\equiv O(t)= U_t^{\dagger}O\sch U_t$, where  $O\sch$ is the corresponding Schr\"odinger operator and $U_{t}=\exp(-i H\sch t/\hbar)$ is the time evolution operator. In this framework, the velocity and the acceleration of the receiver can be expressed respectively as $\dot{X}_1=[X_1,H]/i\hbar$ and $\ddot{X}_1=[\dot{X}_1,H]/i\hbar$. The quantum mechanical work done by the agent on the receiver within a time interval $[t_1,t_2]$ is then defined as \cite{Silva2021}
\begin{equation}
W(t_2,t_1)\!\coloneqq\!\int_{t_1}^{t_2}\!dt\,\,m_1\frac{\{\dot{X}_1,\ddot{X}_1\}}{2}=\Delta K,
\label{RQMW}
\end{equation}
where $\Delta K\!\!=\!K_1(t_2)\!-\!K_1(t_1)$, with $K_1(t)\!=\!\frac{m_1\dot{X}_1^2(t)}{2}$ being the kinetic energy of particle $1$ (receiver's internal energy). Definition~\eqref{RQMW} can be seen as the quantum analog of~\eqref{RQMWcl}, i.e., the Heisenberg statement of the energy-work theorem, which, as shown in Ref.~\cite{Silva2021}, is just a specialization of a more general formulation for quantum systems.

We now proceed to obtain explicit expressions for $W(t_2,t_1)$. Again, we decouple the Hamiltonian operator as $H=P\cm^2/2M+P\rr^2/2\mu+kX\rr^2/2$, by means of the operator transformation
\begin{equation}
\begin{array}{rcl}
X\cm&=&(m_{1}X_{1}+m_{2}X_{2})/M,\\ X\rr&=&X_{2}-X_{1},\\ 
P\cm&=&P_{1}+P_{2},\\
P\rr&=&\mu\left(P_2/m_2-P_1/m_1\right).
\end{array}\label{eq:relacm}  
\end{equation}
As in the classical model, the analytical solutions are very simple and allow us to write \cite{Silva2021} 
\begin{equation}
P_1(t)=a(t)\,P_1\sch+b(t)\,P_2\sch+c(t)\left(X_2\sch-X_1\sch\right),
\label{P1-elastic}
\end{equation}
with the same functions $a(t)$, $b(t)$, and $c(t)$ defined in Eqs.~\eqref{abc}. Restricting again our analysis to the time interval $[t_1,t_2]=[0,v\tau]$, with $\tau=\pi/\omega$ and $v$ an odd integer, and introducing the compact notation $W=W(v\tau,0)$, we find 
\begin{equation}
W=\frac{2}{M^2}(m_1P_2\sch-m_2P_1\sch)(P_1\sch+P_2\sch)
\label{W21-elastic}
\end{equation}
for the operator work done by the agent on the receiver in the process defined by the time interval $[0,v\tau]$. It is clear that $W$ is diagonal in the composite basis $\{\ket{p_1,p_2}\}$, with eigenvalues $w_{p_{1},p_{2}}=\frac{2}{M^2}(m_1p_2-m_2p_1)(p_1+p_2)$ keeping a direct conceptual connection with the classical work~\eqref{W21-elasticl}. Here is the fundamental reason behind our choice of such very particular time intervals; although allowed by our formalism, other choices would demand the numerical diagonalization of $W$. (Possibly, they would also yield different FTs.) Result~\eqref{W21-elastic} tells us that by jointly measuring $P_{1,2}\sch$, one \emph{prepares} an amount $w_{p_1,p_2}$ of work in the interval $[0,v\tau]$. It is worth noticing that one does not really ``measure'' work by measuring the momenta. As discussed in Ref.~\cite{Silva2021} and readily seen from the computations above, a work measurement cannot be performed (not even within the classical paradigm) simply because a two-time observable is not definable at a single time. Instead, we ``prepare'' work for the interval $[0,v\tau]$ through the establishment of $\rho$ at $t=0$.

Having computed the work observable~\eqref{W21-elastic}, we can raise the statistics associated with any well-behaved function $f(W)$ for an initial state $\rho$ acting on the joint space $\mbb{H}$ via
\begin{equation}
\braket{f(W)}_{\rho}=\iint\!dp_1\,dp_2\,f(w_{p_1,p_2})\,\varrho(p_1,p_2),\label{fw}
\end{equation}
where $\varrho(p_1,p_2)=\braket{p_1,p_2|\rho|p_1,p_2}$. In what follows, we analyze the expectation value of the operator $f(W)=e^{-\beta_1 W}$ in instances analogous to those considered in the classical context, but also, and most importantly, in fundamentally quantum scenarios. For the sake of notational compactness and analytical convenience, we introduce the parameters
\begin{equation} 
\epsilon\equiv\frac{\sigma_1}{\Delta_1}\qquad\tx{and}\qquad\gamma\equiv\frac{\sigma_1}{\Delta_1}\frac{\sigma_2}{\Delta_1},
\end{equation}
in terms of which most of the discussion that follows will be conducted. The interpretations of $\sigma_{1,2}$ and $\Delta_1$ are the same ones employed in the classical scenarios of Sec.~\ref{CA Scenarios}.

\subsection{Thermal-Gaussian state}
Let us start with the case involving a thermal-Gaussian state given by
\begin{equation}
\rho_\tx{\tiny TG}=T_{\Delta_1}(\sigma_1)\otimes G_{\bar{\mathbf{r}}_2,\sigma_2},\label{varthgq}
\end{equation}
where $G_{\bar{\mathbf{r}}_2,\sigma_2}=\ket{\bar{\mathbf{r}}_2}\bra{\bar{\mathbf{r}}_2}$ denotes a pure Gaussian state, meaning that
\begin{equation}
\!\!\!\!\!\!\!\!\ba{l}
\braket{x_{2}|\bar{\mathbf{r}}_2}=\left(\frac{2\sigma_2^2}{\pi\hbar^2}\right)^{1/4} e^{-\frac{\sigma_2^2(x_2-\bar{x}_2)^2}{\hbar^2}+\frac{i \bar{p}_2x_2}{\hbar}},\\
\braket{p_{2}|\bar{\mathbf{r}}_2}=\left(2\pi\sigma_2^2\right)^{  -1/4}e^{-\frac{(p_2-\bar{p}_2)^2}{4\sigma_2^2}-\frac{i (p_2-\bar{p}_2)\bar{x}_2}{\hbar}},
\ea
\label{Gp}
\end{equation}
where $\bar{\mathbf{r}}_2=(\bar{x}_2,\bar{p}_2)$ is the centroid and $\sigma_2$ is the agent's momentum uncertainty. For the receiver, we have the effective thermal state
\begin{equation}
T_{\Delta_1}(\sigma_1)=(2\pi\Delta_1^2)^{-1/2}\int\!dp\,e^{-\frac{p^{2}}{2 \Delta_1^2}} \,G_{(0,p),\sigma_1},\label{betast}
\end{equation}
where we recall that $\Delta_1=\sqrt{m_1/\beta_1}$. Being able to avoid the singularities and normalization problems typical of continuum bases and being very convenient for analytical computations, this state actually is an approximation to a genuine thermal state with inverse temperature $\beta_1$. This can be checked from the matrix elements
\begin{equation}
\bra{p_1'}\!T_{\Delta_1}\!(\sigma_1)\!\ket{p_1}
\!=\!\frac{\exp\left[-\frac{{p_1'}^2+p_1^2}{4\Delta_1^2(1+\epsilon^2)}-\frac{\left(p_1'-p_1\right)^2}{8\Delta_1^2\epsilon^2(1+\epsilon^2)}\right]}{\sqrt{2\pi\Delta_1^2 (1+\epsilon^2)}},\nonumber
\end{equation}
which renders, as $\epsilon \to 0$, vanishing coherences and the populations  $\braket{p_1|T_{\Delta_1}(\sigma_1)|p_1}\propto e^{-\beta_1 \mathcal{K}_1}$ with $\mathcal{K}_1=p_1^2/2m_1$. That is, when the momentum fluctuation $\sigma_1$ of the Gaussian state $G_{(0,p),\sigma_1}$  is much smaller than the thermal fluctuation $\Delta_1$, then $T_{\Delta_1}(\sigma_1)$ approaches a fully incoherent mixture (in the kinetic energy basis) with thermal populations. Consequently, whenever $\epsilon\ll 1$, $\rho_\tx{\tiny TG}$ as defined by Eq.~\eqref{varthgq} is a reasonable quantum analog of $\varrho_\tx{\tiny TG}$ as given by Eq.~\eqref{varthg}. Following the prescription indicated by Eq.~\eqref{fw}, with $f(W)=e^{-\beta_1 W}$, we arrive at
\begin{equation}
\!\braket{e^{-\beta_{1} W}}_{\rho_\tx{\tiny TG}}
\!=\!\frac{(m_1+m_2)}{\mf{M}}\exp\left(\frac{2 m_1^2 \epsilon^2 \bar{p}_2^2}{\mf{M}^2\Delta_1^2}\right),
\label{FT-TG}
\end{equation}
where
\begin{equation}
\mf{M}\equiv\sqrt{(m_1-m_2)^2-4m_1(m_1\gamma^2+m_2\epsilon^2)}.
\label{frakM}
\end{equation}
The above FT, which was derived under the convergence condition $\mf{M}^2>0$ for generic values of $\sigma_{1,2}$ and $\Delta_1$, clearly depends on the equilibrium temperature, through $\Delta_1$, but also on the momentum uncertainties $\sigma_{1,2}$ and the masses $m_{1,2}$, whose values regulate the connection with mechanical equilibrium. Now, the series expansion to first order in $\epsilon$, for arbitrary $\sigma_2$, yields
\begin{equation}
\braket{e^{-\beta_1 W}}_{\rho_\tx{\tiny TG}}\cong\frac{m_{1}+m_{2}}{|m_{2}-m_{1}|}\,\Gamma,
\label{FTapp2}
\end{equation}
where $\Gamma\equiv\Big[1-\left(\frac{2m_1\gamma}{m_1-m_2}\right)^2\Big]^{-\frac{1}{2}}$. Note that the instance of a localized agent $(\sigma_2\gg \Delta_1)$ is allowed as long as the convergence condition is preserved. On the other hand, when the agent looses spatial localization, so that $\epsilon\ll\gamma\ll 1$, then we retrieve the classical result~\eqref{FTclthg}, that is, $\braket{e^{-\beta_1 W}}_{\rho_\tx{\tiny TG}}\cong \braket{e^{-\beta_1 \mathcal{W}}}_{\varrho_\tx{\tiny TG}}$, and no dependence on the temperature $\Delta_1$ remains. Again, inertia is seen to play a role in the FT and BK's formula for the nonautonomous scenario is readily retrieved for $m_2\gg m_1$. It is worth emphasizing that the BK context is conceptually approached only when, in addition to $m_2\gg m_1$, we consider a dispersion-free state for the agent. However, this does not guarantee a prescription $\lambda_t$ for the agent's motion, so that, strictly speaking, BK's regime is still not attained.

\subsection{Thermal-thermal state}
In analogy with the classical distribution~ \eqref{varthth}, our next case study focus on the effective thermal-thermal state
\begin{equation}
\rho_\tx{\tiny TT}=T_{\Delta_1}(\sigma_1)\otimes T_{\Delta_2}(\sigma_2),\label{varththq}
\end{equation}
where $T_{\Delta_i}(\sigma_i)$ has the same structure as state~\eqref{betast}. As shown before, when $\sigma_i\ll\Delta_i$ these reduced states become thermal, with respective inverse temperatures $\beta_i$. Direct calculations for generic values of parameters give the exact result
\begin{equation}
\braket{e^{-\beta_1 W}}_{\rho_\tx{\tiny TT}}=	\frac{ (m_1+m_2)}{\sqrt{\mf{M}^2+\epsilon^2 m_1^2(\,\Delta_2/\Delta_1)^2}},
\label{FTthth}
\end{equation}
provided that $\mf{M}^2+\epsilon^2 m_1^2(\Delta_2/\Delta_1)^2>0$. In the regime where $\epsilon\ll\gamma\ll 1$, we obtain the same approximated results of the previous case, so that $\braket{e^{-\beta_1 W}}_{\rho_\tx{\tiny TT}}\cong\braket{e^{-\beta_1 W}}_{\rho_\tx{\tiny TG}}$. However, it is clear that the ratio of temperatures plays a significant role in general.

\subsection{Momentum-momentum correlation state}

Consider the classically correlated quantum state
\begin{equation}
\rho_c=\int\!dp\, \frac{e^{-\frac{ p^{2}}{2 \Delta_1^2}}}{\sqrt{2\pi\Delta_1^2}} G_{(0,p),\sigma}\otimes G_{(0,c p),\sigma}, \label{varcorrthq} 
\end{equation}
with $c\in\mathbbm{R}_{\geq 0}$ being a parameter that correlates the momenta of the particles, in analogy with the scenario defined by Eq.~\eqref{varcorrth}. Notice that here we have $\sigma_2=\sigma_1=\sigma$. Again, directly from prescription~\eqref{fw} we deduce, for generic parameters, that
\begin{equation}
\braket{e^{-\beta_1 W}}_{\rho_c}=	\frac{m_1+m_2}{\sqrt{(m_{1}-m_{2}+2m_{1}c)^2-\mf{F}}}\label{FTcomp}
\end{equation}
under the convergence condition $\mf{F}\leq$ $(m_{1}-m_{2}+2m_{1}c)^2$, where
\begin{equation}
\mf{F}=4 m_1^2 \left[\epsilon^4+\epsilon^2\left(c+\frac{m_2}{m_1}\right)\right].
\end{equation}
Whenever the momentum fluctuations of the Gaussian states are small enough ($\sigma_{1,2}=\sigma \ll \Delta_1$), so that $\epsilon\ll 1$, then the reduced states $\rho_1=\Tr_2 \rho_c$ and $\rho_{2}=\Tr_1 \rho_c$ can be locally identified as thermal, with inverse temperatures $\beta_1$ and $\beta_2=\frac{\beta_1 }{c^2}\frac{m_2}{m_1}$, respectively. In this regime, we find $\mf{F}\cong 0$ and
\begin{equation}
\braket{e^{-\beta_1 W}}_{\rho_c}\cong\frac{m_{1}+m_{2}}{|m_{1}-m_{2}+2m_{1}c|},\label{FTqcorr}
\end{equation}
which agrees with the classical expression~ \eqref{FTclcorr}. Maybe not so surprisingly, in the regime where $\epsilon\ll 1$ we thus far found a complete match between classical and quantum predictions with regard to work FTs within a mechanical perspective. Next, we analyze scenarios without classical counterparts.

\subsection{Agent in quantum superposition}

Let us consider now the preparation
\begin{equation}
\rho_\text{\tiny TS}=T_{\Delta_1}(\sigma_1)\otimes \frac{(\xi+\chi)}{N},\label{varthcq}
\end{equation}
where
\begin{equation}
\ba{r}
\xi=\ket{\bar{\mathbf{r}}_2}\bra{\bar{\mathbf{r}}_2}+\ket{\bar{\mathbf{r}}_2'}\bra{\bar{\mathbf{r}}_2'}, \\
\chi=\ket{\bar{\mathbf{r}}_2}\bra{\bar{\mathbf{r}}_2'}+\ket{\bar{\mathbf{r}}_2'}\bra{\bar{\mathbf{r}}_2},
\label{chixi}
\ea
\end{equation}
with $\ket{\bar{\mathbf{r}}_2}$ and $\ket{\bar{\mathbf{r}}_2'}$ Gaussian states with center at $\bar{\mathbf{r}}_2=(\bar{x}_2,\bar{p}_2)$ and  $\bar{\mathbf{r}}_2'=\bar{\mathbf{r}}_2+(\delta_x,0)$, respectively, and momentum uncertainty $\sigma_2$. $\delta_x$ is a generic spatial displacement and the normalization factor is given $N=\Tr \left(\chi+\xi\right)=2\left[1+\exp\left(-\eta^2/8\right)\right]$, where $\eta\equiv 2\sigma_2\delta_x/\hbar$. In this case, we arrive at
\begin{equation}
\braket{e^{-\beta_1 W}}_{\rho_\text{\tiny TS}}=\braket{e^{-\beta_1 W}}_{\rho_\text{\tiny TG}} \left(\frac{1+e^{-\Omega\,\eta^2}}{1+e^{-\frac{1}{8}\eta^2}}\right)\cos{\Theta},
\label{FTAS}
\end{equation}
where
\begin{equation}
\Omega\equiv 1+\left(\frac{2m_1\gamma}{\mf{M}}\right)^2\!\!\!, \quad \Theta\equiv \frac{\hbar\bar{p}_2}{\delta_x\Delta_1^2} \left(\epsilon\eta\frac{m_1}{\mf{M}} \right)^2\!\!\!,
\end{equation}
again with the convergence condition $\mf{M}^2>0$. Interestingly enough, we see that the ``Gaussian influence'' of the agent's initial state factorizes from the other terms, those which encode via $\eta$ the superposition elements.

\begin{figure}[htb]
\centerline{\includegraphics[scale=0.5]{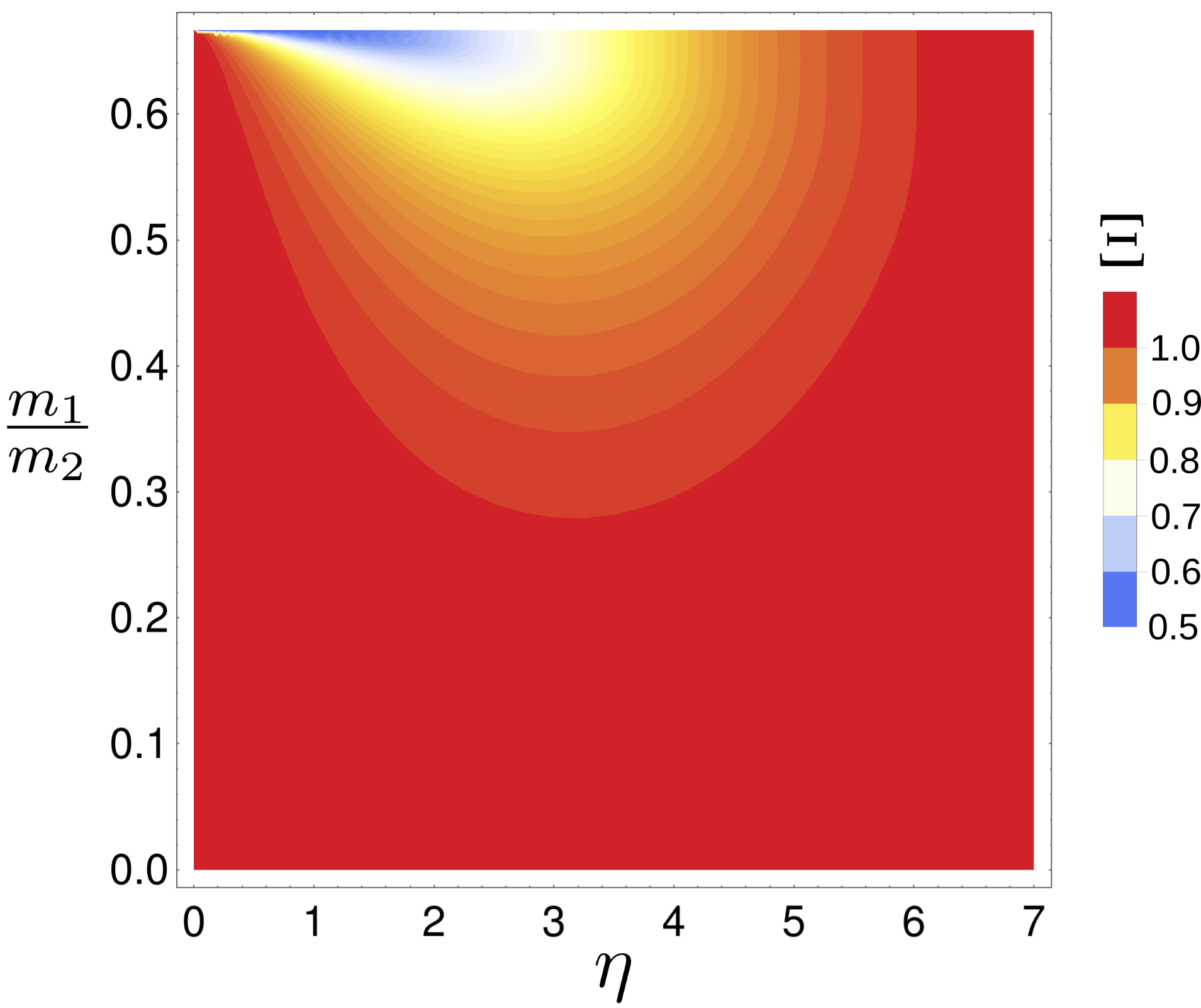}}
\caption{\small Attenuation factor $\Xi$ as a function of the interference parameter $\eta=2\sigma_2\delta_x/\hbar$ and the mass ratio $m_{1}/m_{2}$, in the regime $\sigma_1\ll\Delta_1\ll\sigma_2$ (which is equivalent to $\epsilon\ll \gamma\ll 1$), for $\gamma = 0.25$. The combined effects of inertia and interference are seen to be mostly significant in the upper half plane.}
\label{fig2}
\end{figure}

Different scenarios can emerge from the above result. First, when the local state of the receiver is nearly thermal $(\epsilon\ll 1)$ and no restriction is imposed on the agent initial state, we have $	\braket{e^{-\beta W}}_{\rho_\text{\tiny TS}}\cong	\frac{m_{1}+m_{2}}{|m_{1}-m_{2}|}$, which is no different from the results found when the agent starts in a Gaussian or thermal state. On the other hand, if we consider in addition that $\sigma_2\gg \Delta_1$, so that $\ket{\bar{\mathbf{r}}_2}$ and $\ket{\bar{\mathbf{r}}'_2}$ turn out to be extremely sharp Gaussian states, then we get
\begin{equation}
\braket{e^{-\beta_1 W}}_{\rho_{\text{\tiny TS}}}\cong\frac{(m_1+m_2)}{|m_1-m_2|}\,\Gamma\,\Xi,\label{FTASapp}
\end{equation}
where $\Xi\equiv\frac{1+\exp\left(-\eta^2\Gamma^2/8\right)}{1+\exp\left(-\eta^{2}/8\right)}$. In this case, the role of interference can be analyzed through $\eta$ and $\Xi$. Note that $\eta=2\sigma_2\delta_x/\hbar$ dictates whether a spatial interference pattern is detectable for the preparation: if $\eta\gg 1$, meaning that the distance $\delta_x$ between the wave packets is much greater than their width $\frac{\hbar}{2\sigma_2}$, then no interference pattern is visible via position measurements, although the agent's initial state is a coherent superposition. In this case, one has $\Xi\simeq 1$, and the expression~\eqref{FTASapp} reduces to Eq.~\eqref{FTapp2}, which can be shown to be the FT also when the agent state is prepared in the mixture $\xi/2$. On the other hand, if $\eta$ is not too big, so that interference is observable for the agent's initial state, then $\Xi$ becomes smaller than unit and the FT is significantly influenced by the agent's spatial coherence\footnote{It is worth remarking that the implications of agent's spatial coherence to the work FT cannot be thought of as emerging from local elements solely. To see this, note that if $\gamma=\sigma_1\sigma_2/\Delta_1^2$ (a ``nonlocal'' parameter) could assume vanishing values, then $\Gamma\simeq 1$ and $\Xi\simeq 1$, so that no influence of coherence would survive.}. It can readily be seen from the plot of $\Xi$ as a function of $\eta$ and $m_1/m_2$ (Fig.~\ref{fig2}) how interference and inertia effects can be combined to maximally influence the work FT. Interference becomes most important when, to begin with, it meets the conditions to manifest itself through position measurements on the preparation (which means $\eta$ small) and when the ratio of masses comes closer to its upper bound $(1+2\gamma)^{-1}$, regime which is maximally far apart from the scenario of a heavy agent.

\subsection{Agent entangled with receiver}

At last, we consider an entangled initial state, $\rho_\mf{e}=\ket{\psi_\mf{e}}\bra{\psi_\mf{e}}$, where
\begin{equation}
\ket{\psi_\mf{e}}=\int\!dp\,\frac{e^{-\frac{ p^{2}}{4\Delta_1^2}}}{\sqrt{4\pi\kappa\Delta_1^2}}\ket{0,p}\otimes \ket{\mathfrak{e} p,0},\label{psient}
\end{equation}
$\kappa\equiv\epsilon/\sqrt{1+\epsilon^2+\vartheta_\mf{e}^{-2}}$, $\ket{0,p}$ is a receiver's Gaussian state with center at the mean phase space point $\bar{\mathbf{r}}_1=(0,p)$ and momentum uncertainty $\sigma_1$, and $\ket{\mf{e} p,0}$ is an agent's Gaussian state with center at  $\bar{\mathbf{r}}_2=(\mf{e} p,0)$ and momentum uncertainty $\sigma_2$. The parameter $\mf{e}\in\mathbbm{R}_{\geq 0}$ regulates the correlation of the receiver's momentum with the agent's  position. In the limits $\sigma_1\to 0$ and $\sigma_2\to \infty$, the Gaussian states
$\ket{0,p}$ and $\ket{\mathfrak{e} p,0}$ approach momentum and position eigenstates, respectively, with $\ket{\psi_\mf{e}}$ thus representing a highly entangled state. Moreover, it can be shown that in these limits [as long as $(\sigma_1\sigma_2)^{-1}$ remains bounded] the reduced matrices $\braket{p_1'|\rho_1|p_1}$ and $\braket{x_2'|\rho_2|x_2}$ are nearly diagonal, with respective thermal populations $\exp{\left[-p_1^2/(2\Delta_1^2)\right]}$ and $\exp{\left[-x_2^2/(2\mf{e}^2\Delta_1^2)\right]}$. That is, not only $\rho_1$ is effectively thermal but also $\rho_2$ approximates a mixture of position eigenstates with Gaussian weights of mean value $0$ and dispersion $\mf{e}\Delta_1$.

To make the discussion about quantum correlations quantitative, we compute the amount of entanglement  $E(\rho_\mf{e})=1-\Tr\rho_{\mf{s}}^2$ encoded in $\rho_\mf{e}$ by measuring how far the purity $\Tr\rho_\mf{s}^2$ of the subsystem $\mf{s}\in\{1,2\}$ is from unit (the maximum purity). Direct calculations lead to
\begin{equation}
E(\rho_\mf{e})
= 1- \sqrt{\frac{\epsilon^2\left(1+\epsilon^2+\vartheta_{\mf{e}}^{-2}\right)}{\left(1+\epsilon^2\right) \left(\epsilon^2+\vartheta_{\mf{e}}^{-2}\right)}},
\label{Ent}
\end{equation}
where $\vartheta_{\mathfrak{e}}\equiv\hbar/(2\mathfrak{e}\sigma_1\sigma_2)$. It is straightforward to check that $dE/d\mf{e}\geq 0$, with equality holding for $\mf{e}=0$. This shows that entanglement is a monotonic function of $\mf{e}$, so that, modulo its dimensional unit, this parameter is itself an estimate of entanglement\footnote{While $E(\rho_0)=0$, one has  $E(\rho_\infty)\equiv\lim_{\mf{e}\to\infty}E(\rho_\mf{e})=1-\sqrt{\epsilon^2/(1+\epsilon^2)}$, which does not reach its maximum value when $\mf{e}$ does. However, the definitive monotonicity relation, including maximum and minimum values, can be trivially established, for all $\epsilon$, between $\mf{e}$ and the rescaled measure $E(\rho_\mf{e})/E(\rho_\infty)$.}. It is also interesting to note $\lim_{\epsilon\to\infty}E(\rho_\mf{e})=0$ while  $E(\rho_\mf{e})\cong 1-\epsilon\sqrt{1+\vartheta_{\mf{e}}^2}$ for $\epsilon\ll 1$ (so that $\epsilon=0$ implies maximum entanglement). 

Turning to the FT, in the present case the following exact expression arises:
\begin{equation}
\braket{e^{-\beta_1 W}}_{\rho_\mathfrak{e}}=\frac{m_1+m_2}{\sqrt{\mf{M}^2+\frac{4m_1^2\gamma^2}{1+\vartheta_\mf{e}^{2}(1+\epsilon^2)}}}.\label{FTrhomfe}
\end{equation}
In the regime where $\epsilon\ll 1$, we have
\begin{equation}
\braket{e^{-\beta_1 W}}_{\rho_\mf{e}}\cong\frac{m_{1}+m_{2}}{\sqrt{(m_{1}-m_{2})^{2}-\frac{\hbar^2\beta_1^2}{\mf{e}^2(1+\vartheta_\mf{e}^2)}}}.\label{FTqqcorrapp}
\end{equation}
These results show how entanglement, via $\mf{e}$ and $\vartheta_\mf{e}$, can influence a work FT. It is interesting to assess whether purely classical correlations would also cause a similar impact. To this end, we consider the classically correlated state
\begin{equation}
\rho_\mf{c}=\int\!dp\, \frac{e^{-\frac{ p^{2}}{2 \Delta_1^2}}}{\sqrt{2\pi\Delta_1^2}} G_{(0,p),\sigma_1}\otimes G_{(\mf{c} p,0),\sigma_2},
\end{equation}
which has the same form as $\rho_c$ in Eq.~\eqref{varcorrthq} except that here the momentum of particle $1$ is classically correlated with the position (instead of the momentum) of particle $2$ through the parameter $\mf{c}\in\mathbbm{R}_{\geq 0}$. Just as for $\rho_\mf{e}$, when $\sigma_1\to 0$ and $\sigma_2\to\infty$ with $(\sigma_1\sigma_2)^{-1}$ remaining bounded, the reduced states become nearly thermal. The exact  FT, for arbitrary parameters, turns out to be simply $\braket{e^{-\beta_{1} W}}_{\rho_\mf{c}}=	(m_1+m_2)/\mf{M}$. When $\epsilon\ll 1$ this result can be written as
\begin{equation}
\braket{e^{-\beta_1 W}}_{\rho_\mf{c}}\cong \frac{m_1+m_2}{\sqrt{(m_1-m_2)^2-\frac{\hbar^2\beta_1^2}{\mf{c}^2\vartheta_\mf{c}^2}}},\label{FTqclcorrapp}
\end{equation}
where $\vartheta_\mathfrak{c}\equiv\hbar/(2\mathfrak{c}\sigma_1\sigma_2)$. The formal comparison with result~\eqref{FTqqcorrapp} is now immediate. In particular, we see that the scenarios are comparable when $\vartheta_\mathfrak{e,c}\gg 1$. Also noteworthy is the fact that the relation $\hbar\beta_1/(\mf{c}\vartheta_\mf{c})=2\beta_1\sigma_1\sigma_2$ shows that relation~\eqref{FTqclcorrapp} is $\hbar$-independent and, therefore, can be claimed to be a fundamentally classical result. In any case, though, it is clear that quantum and purely classical correlations, in combination with the thermal and inertial aspects, generally have different impacts on the work FT. This difference disappears as $\hbar\beta_1/m_2$ is sufficiently small, for in this regime both results~\eqref{FTqqcorrapp} and \eqref{FTqclcorrapp} coalesce to the form typically found throughout this article, namely, $(m_{1}+m_2)/|m_{1}-m_{2}|$. Moreover, BK's formula is retrieved as $m_1\ll m_2$. 

Before closing this section, two remarks are in order. First, with regard to energetics, application of Jensen's inequality allows us to write, as in the classical context, $\braket{W}_\rho\geq -\beta_1^{-1}\ln\braket{e^{-\beta_1W}}_\rho$. The state of affairs is then such that, while the BK equality~\eqref{BKintro} imposes that the average work can never be negative in any time interval, here we shown that there exist processes wherein the lower bound for the average mechanical work can assume negative values. This means that, within the present perspective in which work is a Hermitian operator and the system is autonomous, a finite-mass agent can also draw energy from the receiver. Such a result reveals significant deviations from the mechanical equilibrium emerging when $m_2\gg m_1$. 

Second, work FTs are commonly tested by use of two-point measurement (TPM) protocols and incoherent states in the energy basis \cite{Campisi2011,Hanggi2017,Talkner2007}, so it is relevant to examine if and how such methods would deal with the present proposal. Usually, TPM protocols are employed to raise work statistics under the premise that work is a stochastic energy change induced by an external driving parameter $\lambda_t$ \cite{Campisi2011,Hanggi2017,Talkner2007}. This scheme enables the experimental validation of important quantum FTs (see, for instance, Refs. \cite{Campisi2011,Hanggi2017,Huber2008} and references therein) and it gives a relatively simple and fairly general way of accounting for work statistics in the quantum thermodynamics domain. It is often applied to a system $\mr{S}$ described by a time-dependent Hamiltonian $H\sch(t)=H\sch(\lambda_t)$. After being prepared at $t=0$ in a generic state $\rho_\mr{S}$, the system is submitted to a projective measurement of energy at $t_1$, thus jumping to an $H\sch(t_1)$ eigenstate $\ket{e_n}$ with probability $\mf{p}_n=\braket{e_n|\rho_\mr{S}|e_n}$. The system then evolves unitarily (via $U_\dt$, with $\dt=t_2-t_1$) until the instant $t_2$, when a second measurement is performed and a random eigenvalue $\varepsilon_m$ of $H\sch(t_2)$ is obtained with probability $\mf{p}_{m|n}=|\bra{\varepsilon_m} U_{\dt}\ket{e_n}|^2$. In this run of the experiment, work is computed as $w_{mn}=\varepsilon_m-e_n$. After many runs, the work probability density $\wp_w=\sum_{mn}\mf{p}_{m|n}\,\mf{p}_{n}\,\delta_{\tx{\tiny D}}(w-w_{mn})$ is built, where $\int\!dw\,\wp_w=1$. It follows that the $k$-th moment of work can be evaluated as $\overline{w^k} =\int\!dw\,\wp_w w^k= \sum_{mn}\mf{p}_{m|n}\mf{p}_{n}w_{mn}^{k}$. We now examine an adaptation of this protocol to our mechanical perspective. First, it is worth noticing from Eq.~\eqref{P1-elastic} that the kinetic energy operator $K_1$ of particle $1$ at times $0$ and $v\tau$ are such that $[K_1(v\tau),K_1(0)]=0$. Therefore, it might be expected \cite{Talkner2007} that the statistics underlying the work observable $W=K_{1}(v\tau)-K_{1}(0)$ would coincide with TPM predictions. It turns out, however, that this does not materialize for the entangled state $\rho_{\mf{e}}$, since the first measurement of a TPM protocol cancels out the quantum correlation term. To prove this point, we compute the probability density $\wp_{p_{i}}=\left[\left(\ket{p_i}\bra{p_i}\otimes\mathbbm{1}_2\right)\rho_{\mf{e}}\right]$ of finding a momentum $p_i$, and a corresponding kinetic energy $p_i^2/2m_1$, in the first measurement. We find
\begin{equation}
\wp_{p_i}=\frac{\exp\left[-\frac{p_i^2}{2(\Delta_1^2+\sigma_1^2)}\right]}{\sqrt{2\pi (\Delta_1^2+\sigma_1^2)}}.
\end{equation}
As soon as the first measurement is concluded, the state of the system is approximately represented by $G_{(0,p),\sigma_1}\otimes G_{(\mf{c} p,0),\sigma_2}$, with $\sigma_1$ sufficiently small. Now, considering the same procedure for the classically correlated state $\rho_{\mf{c}}$, we find the same probability density for the first measurement, that is, $\Tr \left[\left(\ket{p_i}\bra{p_i}\otimes\mathbbm{1}_2\right)\rho_{\mf{c}}\right]=\wp_{p_i}$. Also, via state reduction, the same state $G_{(0,p),\sigma_1}\otimes G_{(\mf{c} p,0),\sigma_2}$ emerges after the measurement. Therefore, the probability densities related to the first measurement on $\rho_{\mf{e}}$ and $\rho_{\mathfrak{c}}$ are the same and the states right after it also coincide, so that the TPM statistics resulting for $\rho_{\mf{e}}$ and $\rho_{\mf{c}}$ cannot be distinct. Therefore, a relation like Eq.~\eqref{FTrhomfe} cannot be experimentally verified through a TPM protocol, even when the internal energies in the beginning and at the end commute.

\section{Concluding remarks} 

Crucial to the assessment of physical systems' responses to applied perturbations, FTs allow us to analyze averages of fluctuating quantities in terms of physical aspects imposed by thermodynamic equilibrium. Studies in these lines were typically conducted under classical-like assumptions. Nevertheless, searching for the eventual effects of relaxing such constraints is vital for one to build a better comprehension of non-equilibrium thermodynamics, especially in the quantum regime.

In this article, we avoid classicalities in several ways. We consider (i) an exclusive work observable, (ii) a finite-mass agent, (iii) an autonomous agent-receiver dynamics, and (iv) fundamentally quantum global states which are thermal only locally. Then, we compute work FTs for specific processes and proved, by explicit examples, that the BK formula~\eqref{BKintro} cannot be extended to such regimes. Interestingly, we have been able to show that quantum agent's features such as inertia, effective temperature, quantum coherence, and quantum correlations with the receiver directly influence the work FTs. In any case, the BK formula is retrieved for very massive agents, a regime in which energy can only be delivered to the receiver and the dynamics reaches mechanical equilibrium, with the agent in uniform motion. (As shown in the Appendix, though, the recovery of BK's formula can occur in more general scenarios.) Apart from this very particular regime, our FTs show how inertia and quantum resources lead to the breakdown of the mechanical equilibrium. In fact, although our FTs are derived for specific physical processes, we believe that the dependence of FTs on inertia and quantum resources is a fundamental characteristic of any truly autonomous quantum system out of equilibrium. Finally, we show that the usually adopted TPM protocol is unable to capture the influence of entanglement on work FTs.

It is important to stress that the specializations made throughout this article with regard to interactions and time intervals (the physical processes) are motivated merely by analytical convenience. By principle, our mechanical approach to FTs applies to any autonomous quantum system. The key premise is that work is a quantum observable to be computed in the Heisenberg picture; everything else is just standard quantum mechanics. As a consequence, as opposed to traditional results in the field of quantum thermodynamics, the FTs emerging in autonomous contexts are expected to fundamentally depend on the features of each dynamical system under scrutiny.

It would be interesting to explore the work observable formalism of Ref.~\cite{Silva2021} in further autonomous processes and investigating whether some universal bound may eventually pop out. As shown here, the notion of work observable allows us to dig deeper into the extension of FTs to regimes closer to the quantum domain, specially within a fundamentally mechanical perspective. We believe that some of the predictions made here can be experimentally tested in the near future through the promising trapped ion platforms \cite{Rossnagel2016,Levy2020}.

\section{Acknowledgments}
It is a pleasure to thank David Gelbwaser-Klimovsky for comments on a previous version of the manuscript. T.A.B.P.S was supported by the Brazilian funding agency CNPq under Grant No. 146434/2018-8 and the Helen Diller Quantum Center - Technion under Grant No. 86632417. R.M.A. thanks the financial support from the National Institute for Science and Technology of Quantum Information (CNPq, INCT-IQ 465469/2014-0) and the Brazilian funding agency CNPq under Grant No. 309373/2020-4.

%
%
\appendix

\section{Sufficient conditions for the BK limit}
\label{sec:BK_Recover}
Here we identify sufficient conditions enabling the emergence of the BK equality~\eqref{BKintro} out of our work FTs. The analysis will be conducted for both classical and quantum scenarios with the focus on the agent's dynamics.
\subsection{Classical scenario} 
We assume that the receiver $R$ can be any system whose internal energy is described by a classical Hamiltonian  $\mathcal{H}_R(\supdex{\Gamma}{R}_t)$, where $\supdex{\Gamma}{R}_t$ denotes $R$'s phase-space coordinate at time $t$. $\supdex{\Gamma}{R}_t$ can be the position-momentum pair of a single particle, or even any set of canonical variables of a many-particle system. Similarly, we consider a generic agent $A$ with energy $\mathcal{H}_A$ and phase-space coordinate $\supdex{\Gamma}{A}_t$. The composite system $R+A$ evolves in time by means of a Hamiltonian function $\mathcal{H}=\mathcal{H}_R+\mathcal{V}_{RA}+\mathcal{H}_A$, where $\mathcal{V}_{RA}$ is the coupling energy. The joint phase-space coordinate at time $t$ reads $\Gamma_t=\supdex{\Gamma}{R}_t \supdex{\Gamma}{A}_t$.  The work done by $A$ on $R$ during a time interval $[0,t]$ is equal to the variation of $R$'s internal energy~\cite{Silva2021}, that is,
\begin{equation}
\mathcal{W}(t,0)\equiv \mathcal{W}(\Gamma_0,t)=\mathcal{H}_R(\supdex{\Gamma}{R}_t(\Gamma_0,t))-\mathcal{H}_R(\supdex{\Gamma}{R}_0).\label{Wapp}
\end{equation}
Observe that $\supdex{\Gamma}{R}_t=\supdex{\Gamma}{R}_t(\Gamma_0,t)$ is an explicit statement of the fact that, in general, the receiver phase-space coordinate at time $t$ may depend also on the agent's initial condition.

To approach the context underlying the BK's equality, we assume an initial distribution $\varrho(\Gamma_0)$ whose receiver's reduced distribution is thermal, meaning that
\begin{equation}
\varrho_R(\supdex{\Gamma}{R}_0)=\int \!d\supdex{\Gamma}{A}_0\ \varrho(\Gamma_0)=\frac{e^{-\beta \mathcal{H}_R(\supdex{\Gamma}{R}_0)}}{\mathcal{Z}},\label{thermalapp} 
\end{equation}
where $d\supdex{\Gamma}{A}_0$ is the agent's infinitesimal phase-space volume at time $t=0$ and $\mathcal{Z}=\int d\supdex{\Gamma}{R}_0\exp[-\beta \mathcal{H}_{R}(\supdex{\Gamma}{R}_0)]$. To identify the part of $\varrho$ that codifies the correlations between the subsystems, we introduce the function $C(\Gamma_0)=\varrho(\Gamma_0)-\varrho_R(\supdex{\Gamma}{R}_0)\varrho_A(\supdex{\Gamma}{A}_0)$, with $\varrho_A(\supdex{\Gamma}{A}_0)=\int\! d\supdex{\Gamma}{R}_0\,\varrho(\Gamma_0)$. With respect to the generic energy-conserving autonomous dynamics declared so far, the following statement holds.

\begin{result}
The BK equality~\eqref{BKintro} is attained in the aforementioned autonomous scenario whenever the following two conditions are satisfied simultaneously.
\begin{enumerate}
\item The agent's initial phase-space coordinate, $\supdex{\Gamma}{A}_0$, when written in terms of the time $t$ coordinate, does not depend on the receiver's coordinate. Formally, this reads $\supdex{\Gamma}{A}_0\equiv \supdex{\Gamma}{A}_0(\supdex{\Gamma}{A}_t,t)$, where $\supdex{\Gamma}{A}_0(\supdex{\Gamma}{A}_t,t)$ denotes the agent's initial condition that reaches the coordinate $\supdex{\Gamma}{A}_t$ at time $t$.
\item The correlations are such that
\begin{equation}
\int\!d\Gamma_{0}\,e^{-\beta \mathcal{W}(\Gamma_0,t)}\,C(\Gamma_{0})=0.
\end{equation}
\end{enumerate}
\end{result}

\begin{proof}
Assumption 2 implies that 
\begin{equation}
\braket{e^{-\beta \mathcal{W}(t,0)}}=\int\!d\Gamma_{0} \,e^{-\beta \mathcal{W}(\Gamma_0,t)}\,\varrho_R(\supdex{\Gamma}{R}_0)\varrho_A(\supdex{\Gamma}{A}_0).\label{appaux00}
\end{equation}
Using Eqs.~\eqref{Wapp} and \eqref{thermalapp}, along with Hamiltonian invariance of the phase-space volume ($d\Gamma_t=d\Gamma_0$), allows us to rewrite Eq.~\eqref{appaux00} as
\begin{equation}
\braket{e^{-\beta \mathcal{W}(t,0)}}
=\frac{1}{\mathcal{Z}}\int\! d\Gamma_{t}\,e^{-\beta \mathcal{H}_R(\supdex{\Gamma}{R}_t(\Gamma_0,t))}\varrho_A[\supdex{\Gamma}{A}_0(\Gamma_t,t)].\label{auxapp}
\end{equation}
Assumption 1 ensures that $\varrho_A[\supdex{\Gamma}{A}_0(\Gamma_t,t)]=\varrho_A[\supdex{\Gamma}{A}_0(\supdex{\Gamma}{A}_t,t)]$.
Since $d\Gamma_t=d\supdex{\Gamma}{R}_t \,d\supdex{\Gamma}{A}_t$, $\int\! d\supdex{\Gamma}{A}_t\,\varrho_A[\supdex{\Gamma}{A}_0(\supdex{\Gamma}{A}_t,t)]=1$, and
\begin{equation}
\int\! d\Gamma_t^R e^{-\beta \mathcal{H}_{R}[\supdex{\Gamma}{R}_t(\Gamma_0,t)]}=\int\!d\supdex{\Gamma}{R}_0 e^{-\beta \mathcal{H}_R(\supdex{\Gamma}{R}_0)}=\mathcal{Z},
\end{equation}
then $\braket{e^{-\beta\mathcal{W}(t,0)}}=1$ immediately follows.
\end{proof}
The above assumptions reflect two fundamental ingredients through which an autonomous dynamics can approach the regime of deterministic control upon which the BK equality was originally grounded. Assumption 1 indirectly states that the agent's degrees of freedom at a specific time $t$ are independent of the receiver's degrees of freedom at some previous time. This embeds the idea that the agent's dynamics is barely influenced by the receiver. In other words, the agent is allowed to control and deliver energy, but it is by no means controlled. Of course, this regime is also favored when the agent is much heavier than the receiver. By its turn, assumption 2 requires that correlations must not be influential to the work exponential average. Although these are admittedly strong assumptions, they are still weaker than explicitly controlling the agent in a deterministic manner over the entire time interval. Importantly, these assumptions can be materialized in autonomous dynamics. The examples discussed in the main text (for specific time intervals and heavy agents) show that this is indeed the case, that is, the BK equality can emerge out of strictly autonomous dynamics.
%
\subsection{Quantum scenario}
As in the classical case, we consider a generic quantum state $\rho$ acting on a Hilbert space $\mathbbm{H}=\mathbbm{H}_R\otimes \mathbbm{H}_A$, which describes the physics of a system consisting of a receiver $R$ and an agent $A$. Again, the composite system $R+A$ is assumed to evolve autonomously according to a Hamiltonian  $H=H_R+V_{RA}+H_A$, with $H_{R(A)}$ denoting the internal energy of $R\,(A)$ and $V_{RA}$ accounting for the coupling energy. The quantum mechanical work is given by the two-time observable \cite{Silva2021}
\begin{equation}
W(t,0)=H_R(t)-H_R\sch,\label{Wqapp}
\end{equation}
where $H_R(t)=U_t^\dagger H_R\sch U_t$ and $U_t=\exp(-iHt/\hbar)$. Once again, we assume that the receiver's initial state is thermal:
\begin{equation}
\rho_R=\Tr_A\rho=\frac{e^{-\beta H_R\sch}}{\mathcal{Z}},\label{thermalgeneralq} 
\end{equation}
where $\mathcal{Z}=\Tr_R\left(e^{-\beta H_R}\right)$. Following the analogy with the classical case, we define the part $C(\rho)=\rho-\rho_R\otimes \rho_A$ of $\rho$ encompassing the correlations, with $\rho_{R,A}$ being the reduced states of $\rho$. Under these circumstances, we have the following result.
\begin{result}
The BK equality~\eqref{BKintro} is attained in the aforementioned quantum autonomous scenario whenever the following three conditions are satisfied simultaneously.
\begin{enumerate}
\item For any operator $\mathbbm{1}_R\otimes O_A\sch$, it holds that $U_t^\dagger (\mathbbm{1}_R\otimes O_A\sch)U_t=\mathbbm{1}_R\otimes O_A'(t)$ and $U_t (\mathbbm{1}_R\otimes O_A\sch)U_t^\dagger=\mathbbm{1}_R\otimes O_A''(t)$, with $O_A\sch$, $O_A'(t)$, and $O_A''(t)$ acting on $\mathbbm{H}_A$.
\item The correlations are such that
\begin{equation}
\Tr \left[e^{-\beta W(t,0)}C(\rho)\right]=0.
\end{equation}
\item The following relation is valid:
\begin{equation}
\Tr \left\{\left[e^{-\beta W(t,0)}-e^{-\beta H_R(t)}e^{\beta H_R\sch}\right]\rho_R\otimes \rho_A\right\}=0.
\end{equation}
\end{enumerate}
\end{result}
\begin{proof}
From assumption 2, it follows that
\begin{equation}
\braket{e^{-\beta W(t,0)}}=\Tr \left[e^{-\beta W(t,0)}\rho_R\otimes \rho_A\right].\label{appaux1}
\end{equation}
Using assumption 3 and Eqs.~\eqref{Wqapp} and \eqref{thermalgeneralq}, we arrive at
\begin{equation}
\Tr \left[e^{-\beta W(t,0)}\rho_R\otimes \rho_A\right]=\sfrac{1}{\mathcal{Z}}\Tr \left[e^{-\beta H_R(t)}\mathbbm{1}_R\otimes \rho_A\right].\label{appaux2}
\end{equation}
Assumption 1 and the cyclic property of the trace yield
\begin{equation}
\Tr \left[\frac{e^{-\beta H_R(t)}}{\mathcal{Z}}\mathbbm{1}_R\otimes \rho_A\right]=\Tr \left[\frac{e^{-\beta H_R}}{\mathcal{Z}}\mathbbm{1}_R\otimes \rho_A''(t)\right]=1. \label{appaux3}
\end{equation}
Connecting the above equations gives $\braket{e^{-\beta W(t,0)}}=1$.
\end{proof}

In both scenarios, the assumptions 1 and 2 have the same function, namely, implementing the agent's independence, a crucial characteristic of BK's approach. Assumption 3, needed only in the quantum case, circumvents the difficulties underlying the non-commutativity of the algebra. Finally, we remark that the assumptions underlying the results 1 and 2 reflect only sufficient conditions for the recovery of the BK equality. The question of whether these conditions are also necessary, or whether other conditions are relevant, remains to be investigated in future work.


\end{document}